# Interrupted time series analysis of clickbait on worldwide news websites, 2016-2023

Austin McCutcheon
Department of Computer Science
Lakehead University
Orillia, Canada
aomccutc@lakeheadu.ca

Chris Brogly
Department of Computer Science
Lakehead University
Orillia, Canada
cbrogly@lakeheadu.ca

*Abstract*— **Clickbait is deceptive text that can manipulate web browsing, creating an information gap between a link and target page that literally "baits" a user into clicking. Clickbait detection continues to be well studied, but analyses of clickbait overall on the web are limited. A dataset was built consisting of 451,033,388 clickbait scores produced by a clickbait detector which analyzed links and headings on primarily English news pages from the Common Crawl. On this data, 5 segmented regression models were fit on 5 major news events and averaged clickbait scores. COVID and the 2020 US Election appeared to influence clickbait levels.** (*Abstract*)

## I. Introduction

Clickbait and other deceptive-style texts continue to influence web browsing worldwide. Often, clickbait employs sensationalist headlines as link text with the goal of driving user engagement in an attempt to "bait" to target webpages [1]. Furthermore, sometimes the headlines may not accurately reflect what the content truly contains. While the use of clickbait-style links is sometimes fully known and intentional by web users to view different types of content, its underlying deceptive properties are fundamentally questionable in other scenarios. As a result, these styled texts fall into the broader area of deception on the web [1]. Given that clickbait focuses essentially on creating an information gap [1] in order to direct traffic to certain webpages, we were interested in studying how often this style of text might be used by different news sources, and potentially whether or not its use may change as important, reportable world events occur.

Clickbait detection is generally a well-studied problem using natural language processing (NLP) [1], machine learning (ML) [1], [2] and deep learning (DL) techniques [3]. NLP and ML/DL techniques have shown to be very accurate [2] ways of detecting clickbait headlines given sufficient train/test datasets. Overall, there has been more focus on building detectors using these methods and reporting on performance metrics with the relevant clickbait datasets, with less focus on how we can use these tools to further understand clickbait trends at scale. One potential benefit of investigating clickbait at scale is that it may become possible to observe trends related to when it occurs more. For instance, reportable world events could influence the production of clickbait on news websites, where the text style is sometimes used in contrast to the more standard headlinese. By applying an existing clickbait detector on a crawl of the web, a clickbait score can be assigned to each hyperlink on a webpage. These scores can be aggregated from a large number of pages and further analyzed to determine if any associations might exist.

Considering the potential for world events to generate news, which may influence the level of deceptive text online, we were interested in analyzing levels of clickbait with respect to significant world events over the last few years in online news websites. Given that clickbait is designed to create an information gap, and, in general, news webpages are meant to provide information, it is not well understood how a deceptive style of text like clickbait could be associated with key content-generating events online. As a result, our goals were to 1) build a dataset of clickbait scores assigned to all links on primarily English-language news webpages with an established clickbait detector, and 2) create statistical models using interrupted time series analysis that might relate levels of clickbait to a selected number of world events to determine if any associations might exist. The following 5 world events were selected for the statistical models: 1) US Election 2016, 2) COVID Public Health Emergency of International Concern (PHEIC), 3) COVID Pandemic, 4) US Election 2020, and 5) the launch of ChatGPT. These events were selected for this work due to their fundamental aspects being completed or primarily completed by the time of writing, and also that they were widely reported on globally. These were the only criteria for selected world events.

## II. Methods

### A. Worldwide news website hyperlink-clickbait score dataset development

First, a dataset of hyperlink text with associated clickbait scores was created. This was derived from the Common Crawl news webpages dataset ranging from 2016/09/02 to 2023/06/28 (data beyond this date was not publicly available when we examined it). Custom software was developed to parse the HTML pages every Tuesday and Friday from the Common Crawl news dataset and extract all hyperlink captions in order to pass those into a clickbait detector that produced a prediction as to whether or not the text may be clickbait. The clickbait detector that was used for this work is an upgraded version of a clickbait detector developed previously by Brogly and Rubin [1]. The only difference between the version used here and the original in [1] is that the clickbait training data was updated to more recent examples of the style of clickbait text using a newer dataset [4], and a now unsupported NLP library was replaced with spaCy. This detector performs with 93% accuracy on a test clickbait dataset, is well-documented with associated publications [1], [5], and has been used to analyze clickbait in a graphical program that underwent peer-review [5]. Furthermore, as it only uses standard machine learning, the scores for each feature that the detector uses to produce a classification result of either 1) clickbait or 2) not clickbait (or the associated scores with a range of 0.0-1.0) could be saved as part of our database. The resulting database has 7 populated columns: id (autoincrementing ID of a potential clickbait text), ymd (year-month-day timestamp), tag (HTML tag, which could include a, h1-h5), pageurl (url of the page this link came from), headline (text of potential clickbait headline), detector (comma-separated values of the detector machine learning features), score 1 (clickbait prediction, 0.0-1.0), and score 2 (not-clickbait prediction, 0.0-1.0, or 1.0-clickbait prediction). The dataset is saved in standard SQLite database files, with a total size of 168GB consisting of 451,033,388 rows that include the above mentioned columns.

*B. Interrupted time series analysis*

For all statistical analysis, R version 4.4.0 was used, along with the latest lmtest, car, and nlme packages. After building the database using the Common Crawl, the 451,033,388 clickbait scores were averaged to 708 total data points, one representing each unique day a page was crawled in our dataset. This was done to simplify running statistical models on the data as the original N=451,033,388 took significantly longer to process. Following this, an interrupted time series analysis was performed between the clickbait scores and each of the selected world events. Initially we fit 5 standard R lm() models with variables relevant to an interrupted time series, and then assessed ACF and PACF results (all using a lag of 20) to determine if autocorrelation was an issue in the data, which it was. This was compensated for with a correlation structure of ARMA(p=16, q=0), resulting in models with lower AIC values. These 5 final generalized least squares (GLS) regression models were fit – and are reported on in the following section - to determine if there were any significant associations with the events and the clickbait scores.

## III. RESULTS

From the dataset, only the average clickbait score per day and the day itself were used to complete the time series analysis; other columns were disused. Key information regarding the basic properties of the dataset used, including summary statistics and collection parameters (for example minimum word count of processed text) are shown below in Table 1.

A histogram of all 451, 033, 388 headlines was plotted. It is below in Fig. 1. It is right skewed, showing lower clickbait scores, suggesting more use of headlinese from news sites.

TABLE 1: DATASET SUMMARY STATISTICS/PROPERTIES

| Dataset Property | Value |
|---|---|
| Total number of unique news websites analyzed | 26212 |
| HTML tags processed | "a", "span", "h1", "h2", "h3", "h4", "h5", "yt-formatted-string" |
| Hyperlink captions/headings analyzed (sample size/N) | 451,033,388 (also see Fig. 1) |
| Minimum word requirement for processed text | 3 |
| Mean clickbait score | 0.32729 |
| Median clickbait score | 0.2566 |
| Mode clickbait score | 0.38838 |
| Days with clickbait averaged and used for time series | 708 |
| Number of news-relevant events selected for interrupted time series | 5 |
| Significant p-values with Bonferroni correction | < 0.01 (0.05 / 5) |

FIG. 1: HISTOGRAM OF ALL TEXT CLICKBAIT SCORES

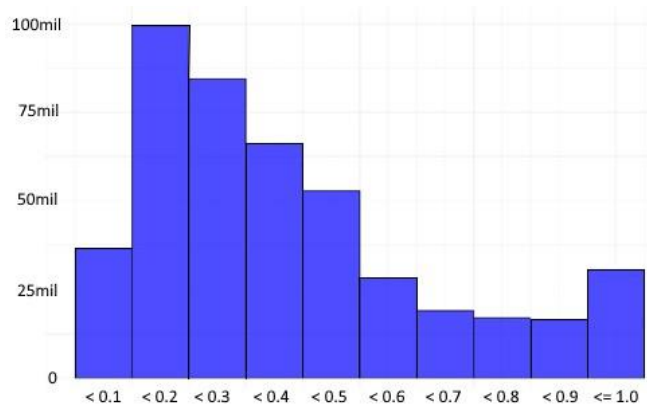

Since there were 5 world events being tested on the same dataset, p < 0.05 with Bonferroni correction applied resulted in p < 0.01 being considered significant. For an interrupted time series, in order to fit our final GLS regression models, three variables were added to the daily average clickbait scores. These needed variables are used from and described in [6], and are as follows:

- **Time Trend (T):** Was there a statistically significant increase or decrease in clickbait every day before the event occurred?

- **Event Impact (D):** Was there a statistically significant immediate increase or decrease in clickbait score when the world event occurred?

- **Post-Event Trend (P):** Have clickbait levels changed after the occurrence of the world event? For each day that occurs after the event, is there a sustained and significant increase or decrease in clickbait level?

TABLE 2: GLS SEGMENTED REGRESSION TIME SERIES FITS

| US Election 2016, Nov. 8, 2016 AIC = -4258.903, ARMA(p=16, q=0) | | | | | |
|---|---|---|---|---|---|
| | *Value* | *Std.Error* | *t-value* | *p-value* | *Sig?* |
| T | -0.000014 | 0.000209 | -0.0701 | 0.9441 | N |
| D | -0.000148 | 0.008649 | -0.0171 | 0.9863 | N |
| P | 0.000010 | 0.000209 | 0.0511 | 0.9592 | N |
| **COVID-19 WHO PHEIC Declaration, Jan. 30, 2020 AIC = -4267.353, ARMA(p=16, q=0)** | | | | | |
| | *Value* | *Std. Error* | *t-value* | *p-value* | *Sig?* |
| T | -0.000017 | 0.000006 | -2.5897 | 0.0098 | Y |
| D | -0.003721 | 0.005701 | -0.6527 | 0.5141 | N |
| P | 0.000030 | 0.000010 | 2.8847 | 0.0040 | Y |
| **COVID-19 WHO Pandemic Declaration March 11, 2020 AIC = -4267.011, ARMA(p=16, q=0)** | | | | | |
| | *Value* | *Std.Error* | *t-value* | *p-value* | *Sig?* |
| T | -0.000019 | 0.000006 | -3.0179 | 0.0026 | Y |
| D | 0.000993 | 0.005695 | 0.1744 | 0.8616 | N |
| P | 0.000030 | 0.000010 | 2.8740 | 0.0042 | Y |
| **US Election 2020, Nov. 3, 2020 AIC = -4267.647, ARMA(p=16, q=0)** | | | | | |
| | *Value* | *Std.Error* | *t-value* | *p-value* | *Sig?* |
| T | -0.000014 | 0.000005 | -2.8147 | 0.0050 | Y |
| D | -0.003606 | 0.005719 | -0.6305 | 0.5285 | N |
| P | 0.000034 | 0.000011 | 3.0215 | 0.0026 | Y |
| **Launch of ChatGPT, Nov. 30, 2022 AIC = -4259.434, ARMA(p=16, q=0)** | | | | | |
| | *Value* | *Std. Error* | *t-value* | *p-value* | *Sig?* |
| T | -0.0000054 | 0.000004 | -1.1888 | 0.2349 | N |
| D | -0.0008877 | 0.007111 | -0.1248 | 0.9007 | N |
| P | 0.0000440 | 0.000062 | 0.7010 | 0.4835 | N |

## IV. DISCUSSION

In terms of significant results, the coefficients from these models are small, although this was anticipated since the dataset consisted of only daily averages of clickbait in the range of 0.0-1.0. The models for both COVID events and the 2020 election produced some significant results. There appeared to be a slight decrease in clickbait each day towards these events, with COVID's pandemic declaration being the largest decrease. None of the models showed a significant change in clickbait score immediately when the event occurred. For each day after these events, there is a sustained and slight increase in clickbait level. The coefficient is small, but significant. The COVID PHEIC and Pandemic event sustained change is identical (the events are related with the second being an escalation of the first) and we see slightly more after the 2020 US Election.

The model for the 2016 US Election since not produce any significant terms. There are some suspected reasons for this. First, there was only about 2 months of data before the 2016 election day, which was much less than in our other models, which may have impacted the segmented regression fit. It is also possible that a significant change in clickbait started occurring well in advance of this election due to widespread media coverage, which may be a reason that the model did not produce any significance. Since we only had data from mostly the fourth quarter of 2016, we cannot look at any dates earlier than election day.

The model for the launch of ChatGPT did not produce any significant terms either. Although it may seem unrelated to COVID and election events, we felt, as with the other choices, it was potentially an important news-impacting event with respect to clickbait given the ability to produce realistic stories and other text content using this large language model (LLM). One limitation from this model fit, similar to the US election 2016, is that there was less data after the introduction of ChatGPT (7 months only) in comparison to COVID and the 2020 election (2.5 years), although this was not as limited as, for instance, the 2 months available prior to the 2016 US election. Still, this may have impacted the results. However, the non-significance of the model may suggest that ChatGPT generated text was not being widely deployed on news sites from our data, or, at least that the headlines/links it generated do not seem to follow the general style of clickbait text. Further research directly using ChatGPT or comparable advanced LLMs is needed to add support to the findings here (or to contradict them).

## V. LIMITATIONS

There are limitations to this work. In terms of our dataset, the individual clickbait scores of the sample of 451 million headlines were averaged to the 708 crawl dates that were available; this was done in order to reduce processing time of the data at the potential expense of some statistical information. In terms of the stationarity of our dataset, an ADF test suggested stationarity, however, a KPSS test did not. Another limitation is that it is difficult to control for confounding variables in our segmented regression models. While we feel our analysis is reasonable given that we selected news-influencing events and a low p-value threshold, it is still possible there are other factors influencing these results that we could not control for. Furthermore, this research is predicated on the assumption that our clickbait detector usually would produce accurate ratings of previously unseen text. We openly admit there will be error in the dataset as we already know the detector does not perform at 100% in all cases. Even a detector at 99.9% accuracy would likely not perform at that level on our sample size since there will be a significant amount of text outside of previously seen training examples. There are potential biases in the data due to taking data primarily from English websites. The minimum word requirement of 3 in a headline allowed some webpage structural links, and others into the dataset that did not need any analysis. However, it is not always clear what text should be considered as possible clickbait and what should not be. Additionally, our English-detecting function was fast but basic, so it did sometimes allow other languages that used the same characters into the dataset.

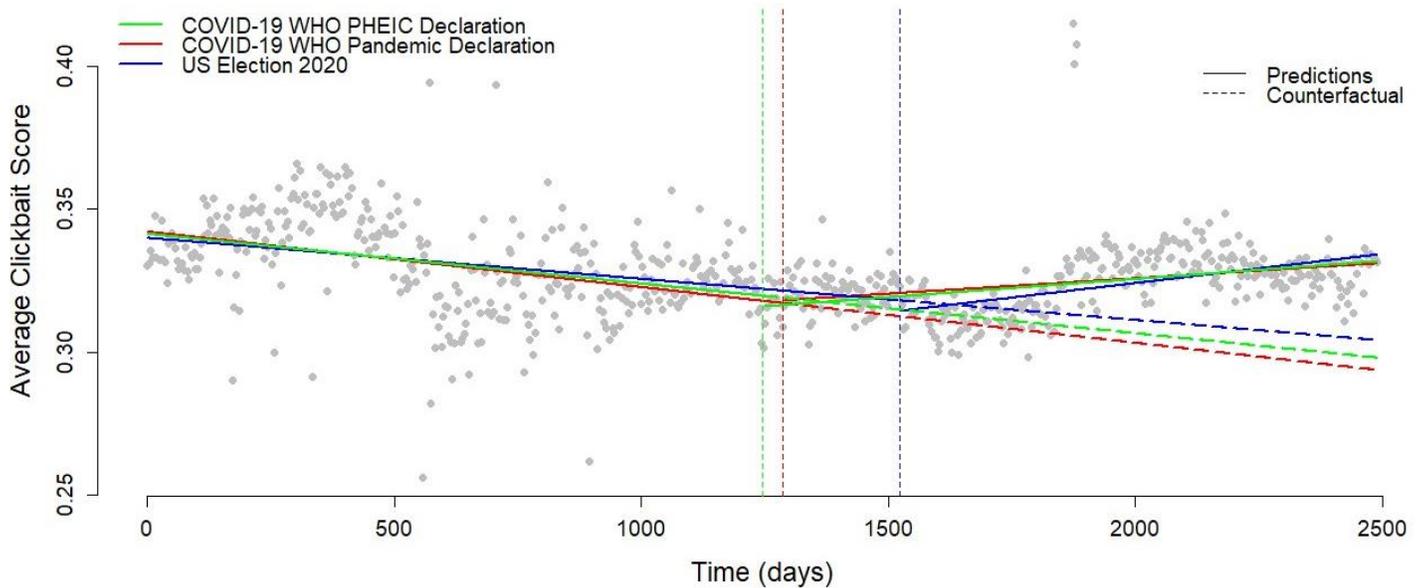

FIG. 2: PLOT OF ALL SEGMENTED REGRESSION MODELS WITH SIGNIFICANT TERMS

## VI. CONCLUSION

This research provides an analysis of clickbait at the scale of the web on primarily English-language news sites from the Common Crawl. In this analysis, 3 of 5 selected news-generating events produced significant terms that suggest major events can impact clickbait levels. COVID declared as both a public emergency and a pandemic seemed to result in a sustained increase in clickbait after these events. The 2020 US Election model showed this as well, although the coefficient is only slightly higher than the COVID models. While segmented regression was used here and is commonly found in ITS studies, there are other models applicable to this data as well.

This analysis should only be considered as one perspective on clickbait with respect to our 5 chosen world events. There were a number of stated limitations, primarily arising from the fact that applying any sort of statistical learning predictions on large datasets from the web will introduce error and make controlling for all influencing factors difficult. Even considering this, understanding a style of text like clickbait across the broader web should provide insights into the phenomenon. There remain opportunities to explore clickbait at a larger scale and also in contexts outside of news. The dataset used in this paper will be publicly available on Zenodo (URL goes here for print copy) for other interested researchers to use.

## ACKNOWLEDGMENT

This research was supported by a grant from Lakehead University, Faculty of Science and Environmental Studies. C.B. thanks Sarah Paleczny (IPRO, St. Michael's Hospital) for input on time series analysis.